\documentclass{aa}
\usepackage{natbib}
\bibpunct{(}{)}{;}{a}{}{,} 
\usepackage{graphicx}
\usepackage{amsmath}
\usepackage{gensymb}
\usepackage{multirow}
\usepackage{subcaption}
\usepackage{txfonts}
\usepackage[normalem]{ulem}

\newcommand{\Msol}{M_\odot}

\usepackage[colorinlistoftodos,prependcaption,textsize=tiny]{todonotes}

\usepackage{natbib}

\begin{document} 

\title{
New limits from microlensing on Galactic Black Holes in the mass range $10\Msol<M<1000\Msol$
}
\author{
T.~Blaineau\inst{1},
M.~Moniez\inst{1},
C. Afonso\inst{2},
J.-N. Albert\inst{1},
R. Ansari\inst{1},
E. Aubourg\inst{2,3},
C. Coutures\inst{2},
J.-F. Glicenstein\inst{2},
B. Goldman\inst{4,5},
C. Hamadache\inst{1},
T. Lasserre\inst{2},
L. Le Guillou\inst{6},
E. Lesquoy\inst{2},
C. Magneville\inst{2},
J.-B. Marquette\inst{7},
N. Palanque-Delabrouille\inst{2,8},
O. Perdereau\inst{1},
J. Rich\inst{2},
M. Spiro\inst{2},
P. Tisserand\inst{9}
}
\institute{
Laboratoire de physique des 2 infinis Ir\`ene Joliot-Curie,
CNRS Universit\'e Paris-Saclay,
B\^at. 100, Facult\'e des sciences, F-91405 Orsay Cedex, France
\and
IRFU, CEA, Universite de Paris-Saclay, F91191 Gif-sur-Yvette, France
\and
Université de Paris, CNRS, Astroparticule et Cosmologie,  F-75013 Paris, France
\and
The International Space University, 1 Rue Jean-Dominique Cassini, F-67400 Illkirch-Graffenstaden, Germany
\and
Université de Strasbourg, CNRS, Observatoire Astronomique, CNRS, UMR 7550,F-67000, Strasbourg, France
\and
Sorbonne Universit\'e, CNRS/IN2P3,
Laboratoire de Physique Nucl\'eaire et de Hautes \'Energies (LPNHE),
75005 Paris, France
\and
Laboratoire d'astrophysique de Bordeaux, Univ. Bordeaux, CNRS, B18N, allée Geoffroy Saint-Hilaire, 33615 Pessac, France
\and
Lawrence Berkeley National Laboratory, One Cyclotron Road, Berkeley, CA 94720, USA
\and
Sorbonne Universit\'es, UPMC Univ Paris 6 et CNRS, UMR 7095, Institut d'Astrophysique de Paris, IAP, F-75014 Paris, France
}

\offprints{M. Moniez,\\ \email{ moniez@lal.in2p3.fr}}

\date{Received 28/02/2022, accepted 3/06/2022}
%

\abstract
{
We have searched for long duration microlensing events originating from intermediate mass Black Holes (BH) in the halo of the Milky Way, using archival data from EROS-2 and MACHO photometric surveys towards the Large Magellanic Cloud.
We combined data from these two surveys to create a common database of light curves for 14.1 million objects in LMC, covering a total duration of 10.6 years, with flux series measured through four wide passbands. We have carried out a microlensing search on these light curves, complemented by the light curves of 22.7 million objects, observed by EROS-2 only or MACHO only over about 7 years, with flux series measured through only two passbands.
A likelihood analysis, taking into account LMC self lensing and Milky Way disk contributions allows us to conclude that compact objects with masses in the range $10 - 100 \Msol$ cannot make up more than $\sim 15\%$ of a standard halo total mass (at $95\%$ confidence level). Our analysis sensitivity weakens for heavier objects, although we still exclude that $\sim 50\%$ of the halo be made of $\sim 1000 \Msol$ BHs.
Combined with previous EROS results, an upper limit of $\sim 15\%$ of the total halo mass can be obtained for the contribution of compact halo objects in the mass range $10^{-6} - 10^2 \Msol$.
} 


\titlerunning{New limits from microlensing on Galactic Black Holes}
\authorrunning{Blaineau T., Moniez M.}

\maketitle

\section{Introduction}
Observation of gravitational waves due to coalescence of massive objects \citep{Abbott_2016b, Abbott_2016a} have demonstrated the existence of merging black holes heavier than $10 M_{\odot}$, and has renewed interest in Black Holes as dark matter, especially primordial Black Holes (PBH) \citep{Bird_2016,2021JPhG...48d3001G,2016PhRvL.117f1101S}.
Microlensing surveys toward the LMC,
that allow to probe the content of the Galactic halo in massive compact objects,
have shown that objects lighter than $10 M_{\odot}$ do not significantly contribute to the hidden mass of the Galactic spherical halo of our galaxy \citep{Tisserand_2007,MACHO_2001,Wyrzykowski_2011}.
Since the typical duration of microlensing events increases with the lens mass, the detection of heavier objects such as those responsible for gravitational wave emissions needs time-series data longer than the durations of each of the EROS-2 (Exp\'erience de Recherche d'Objets Sombres) and MACHO (MAssive Compact Halo Objects) surveys, two of the first microlensing surveys that operated in the years 90s and 2000s.
To explore the dark matter halo beyond this limit by searching for events with longer timescales, we have combined the databases of EROS-2 and MACHO,
thus starting the program described in \citet{Mirhosseini2018}.

In Sect. \ref{section:lensingLMC} we describe the microlensing effect, focusing on the LMC searches.
Section \ref{section:combine} introduces the EROS-2 and MACHO surveys and their light curve datasets, summarizes
how we associated objects between the two catalogs, and presents procedures that allowed us to remove defective images and measurements.
In Sect. \ref{section:selection} we describe the selection of candidates for gravitational microlensing events, based on fitting the observed light curves with theoretical microlensing curves.
In Sect. \ref{section:efficiency} we explain the calculation of the efficiency for detection of microlensing events.
We quantify the effect of blending on the detection efficiency by using HST data, and discuss the impact of the binary sources.
In Sect. \ref{section:resultats},
we confront the number of selected candidates with the number of events expected from the dark matter halo, from the Galactic disk, and from the LMC itself. We then derive a
new upper limit on the contribution of compact objects to the halo.
Finally, in Sect. \ref{section:discussion} we list the sources of improvement we achieved with respect to previous results, and propose some perspectives for further data combinations.

\section{Microlensing toward LMC}
\label{section:lensingLMC}
A gravitational microlensing effect occurs when a massive compact object (called a lens or a deflector in the following) passes close enough to the line of sight of a star to produce gravitational images that are not intercepted by the lens.
The size of the opaque part of the lens and the relative positions of the source, lens and observer must be such that the rays of the two gravitational images are not occulted (no eclipse). For typical lens-source configurations considered here, the angular separation of the two images is too small to be resolved in telescopes. The detection of the event is made possible by the relative motion of the lens in the observer-source frame, which produces a transient variation of the source brightness.

Microlensing of Large Magellanic Cloud (LMC) stars as a technique to search for massive compact objects in the Galactic halo was first described in \citet{Paczynski86}. Reviews of the formalism can be found in \citet{Schneider_2006} and \citet{Rahvar_2015}.
When a point-like object (lens) of mass $M_L$ located at distance $D_L$ passes
close to the line of sight of a point source located at distance $D_{LMC}=49.5 kpc$ \citep{Pietrzynski2019}, the total magnification of the source luminosity at a given time $t$ is the sum of the contributions of two images, given by:
\begin{equation}
A(t)=\frac{u(t)^2+2}{u(t)\sqrt{u(t)^2+4}},
\label{Amplification}
\end{equation}
where $u(t)$ is the distance of the lens to the undeflected line of sight,
divided by the Einstein radius $r_{\mathrm{E}}$,
\begin{equation}
r_{\mathrm{E}}\! =\!\! \sqrt{\frac{4GM_L}{c^2}D_{LMC} x(1\! -\! x)}\!
\simeq\! 10.0\mathrm{AU}\times\left[\frac{M_L}{\Msol}\right]^{\frac{1}{2}}\!
\frac{\left[x(1\! -\! x)\right]^{\frac{1}{2}}}{0.5}\!. 
\end{equation}
Here $G$ is the Newtonian gravitational constant and $x$ the lens to source distance ratio \mbox{$x = D_L/D_{LMC}$}.
The Einstein radius of the lens is such that a point source that is behind the Einstein disk (of surface $\pi r_{\mathrm{E}}^2$), sees its apparent luminosity magnified by a factor greater than 1.34.
Assuming that the lens has a constant relative transverse velocity $v_T$,
$u(t)$ is given by:
\begin{equation}
u(t)=\sqrt{u_0^2+(t-t_0)^2/t_{\mathrm{E}}^2},
\end{equation}
where $t_{\mathrm{E}}=r_{\mathrm{E}} /v_T$ is the Einstein radius crossing time,
and $u_0$ is the minimum distance to the undeflected line of sight at time $t_0$.

In the approximation of a point lens acting on a point source, with a uniform relative motion with respect to the line of sight (hereafter called PSPL approximation), the microlensing effect has several characteristic
features which allow one to discriminate it from any known
intrinsic stellar variability:
\begin{itemize}
\item {
Given the low probability of alignment required for a measurable microlensing effect to occur, it is expected that such an event will not be repeated for a given source or lens on typical human time scales.}
\item The magnification is a known function of
time, depending on only 3 parameters ($u_0, t_0, t_{\mathrm{E}}$),
with a symmetrical shape, and independent of the passband.
\item Since the geometric configuration of the source-deflector system
is random, the prior distributions of $t_0$ and of the impact parameters $u_0$ of the events must be uniform.
\item The passive role of the lensed stars implies that they
should be representative of the monitored sample.
\end{itemize}

If the gravitational field of the Galaxy is entirely due to the lenses, the optical depth for lensing, {\it i.e.} the probability that the line of sight to an LMC star is within one $r_E$ of a lens is of order $v_{rot}^2/c^2$, where $v_{rot}$ is the orbital velocity around the Galaxy at the LMC position.
We use as a benchmark model the isotropic and isothermal halo first
studied by \citet{1991ApJ...366..412G} (hereafter called S-model), with the mass density distribution
\begin{equation}
\rho_{H}(r) = 0.0078 \frac{R_{0}^{2}+R_{c}^{2}}{r^{2}+R_{c}^{2}}\Msol pc^{-3}\ ,
\end{equation}
where
$R_{0}=8.5\,$kpc is the Galactocentric distance to the Sun,
$R_{c}=5\,$kpc the halo ``core radius'' and $r$ the Galactocentric radius.
We use this model with the most recent values of the LMC distance $D_{LMC}=49.5\,$kpc 
\citep{Pietrzynski2019}.

The velocity $v_L$ of the halo objects follows the Maxwell distribution
\begin{equation}
p(v_L)=\left(\frac{1}{2\pi v_0^2}\right)^{3/2}4\pi v_L^2 e^{-v_L^2/2v_0^2},
\end{equation}
with $v_0=120\,$km$\,$s$^{-1}$;
the velocity of the Sun is $\bf{v_{\odot}}$ $=(11.1,251,7.3)\,$ km$\,$s$^{-1}$ \citep{Brunthaler_2011} in galactocentric Cartesian coordinates (X pointing from the Sun to the Galactic center, Z pointing north),
and the proper velocity of the LMC is
$\bf{v}_{LMC}$ $=(-57,-226,221)\,$ km$\,$s$^{-1}$
\citep{Kallivayalil_2013}.

This model gives an
optical depth to the LMC  $\tau_{LMC}\sim4.7\times 10^{-7}$
if the halo is completely made of compact objects.
The total rate (events per star per unit time) 
for $u_0<1$ is $\Gamma=(2/\pi)\tau_{LMC}/\langle t_E\rangle$ \citep{1991ApJ...366..412G},
where
$\langle t_E \rangle$ is the mean $t_E$ of all events
\footnote{ 
\citet{Rahvar_2015} uses $\langle 1/t_E \rangle$ instead of $1/\langle t_E\rangle$ to estimate the event rate. The former refers to an average over all events in progress (within an Einstein ring) at a given time, whereas the latter refers to an average over all events occurring within a given time interval.
}.
For lenses of mass $M_L$,
the  mean event duration  for this benchmark halo model
is $\langle t_E\rangle\sim 63\,day \sqrt{M_L/\Msol}$.

\section{Combining EROS-2 and MACHO data}
\label{section:combine}
The EROS-2  and MACHO surveys were performed  with similar setups, respectively installed at the La Silla Observatory (ESO, Chile) and at the Mount Stromlo Observatory (Australia) (see table \ref{tab:surveys}).
EROS-2 used of a 1 meter (F/5) diameter telescope, equipped with a dichroic beam splitter and two cameras, each with 8 2Kx2K CCD's, covering $1.\,$deg$^2$.
MACHO used a slightly larger telescope 
($1.27m$, F/3.9), equipped with two cameras, each with 4 2Kx2K CCD's, covering a smaller field of view ($0.5\,$deg$^2$).
\begin{table}
    \begin{center}
\begin{tabular}{ccc} \hline \\  [-1ex]
 & EROS-2 & MACHO \\
\hline
telescope & 1 m & 1.27 m \\
pixel size & $0.62"$ & $0.63"$ \\
blue passband & $[420,720]\,$nm & $[450,590]\,$nm  \\
red passband & $[620,920]\,$nm & $[590,780]\,$nm \\
median image quality & $2"$ & $2.1"$ \\
 \hline
\end{tabular}
\centering
\caption[]{Characteristics of EROS-2 and MACHO setups.
}
\label{tab:surveys}
\end{center}
\end{table}

EROS-2 surveyed 88 fields ($1.\,$deg$^2$) toward LMC, and MACHO  surveyed 82 ($0.5\,$deg$^2$) LMC fields.
MACHO adopted the same exposure time (300s) for all fields, whereas
EROS-2 has chosen to adapt its exposure times according to the surface brightness of the fields, varying from 180s (for the central fields) to 900s (for the external fields). From the end of 1999, the longest exposure times were reduced in EROS-2, in order to increase the overall sampling of the survey.
The EROS-2 object catalog was produced after co-adding at least 10 images per field, and rejecting the faintest and brightest objects.
This allowed it to partially compensate for the effect of a shorter exposure in the densest fields and a smaller telescope diameter compared to MACHO.

The MACHO light curves and images are publicly available\footnote{https://macho.nci.org.au/} \citep{Alcock1999}. The catalog contains $22.3\times10^6$ objects with magnitudes $V<21.5$, of which $6\%$ are duplicate objects due to overlapping fields.
The light curves cover a duration of 7.7 years, longer than the duration analyzed in the last MACHO publication (5.7 years) \citep{MACHO_2001}.

The EROS-2 catalog for LMC was produced by \citet{TisserandThese} for the final EROS LMC publication \citep{Tisserand_2007}.  It will  be made public  to allow future expansion of the work described in this paper.

\begin{figure}
    \centering
    \includegraphics[width=\linewidth]{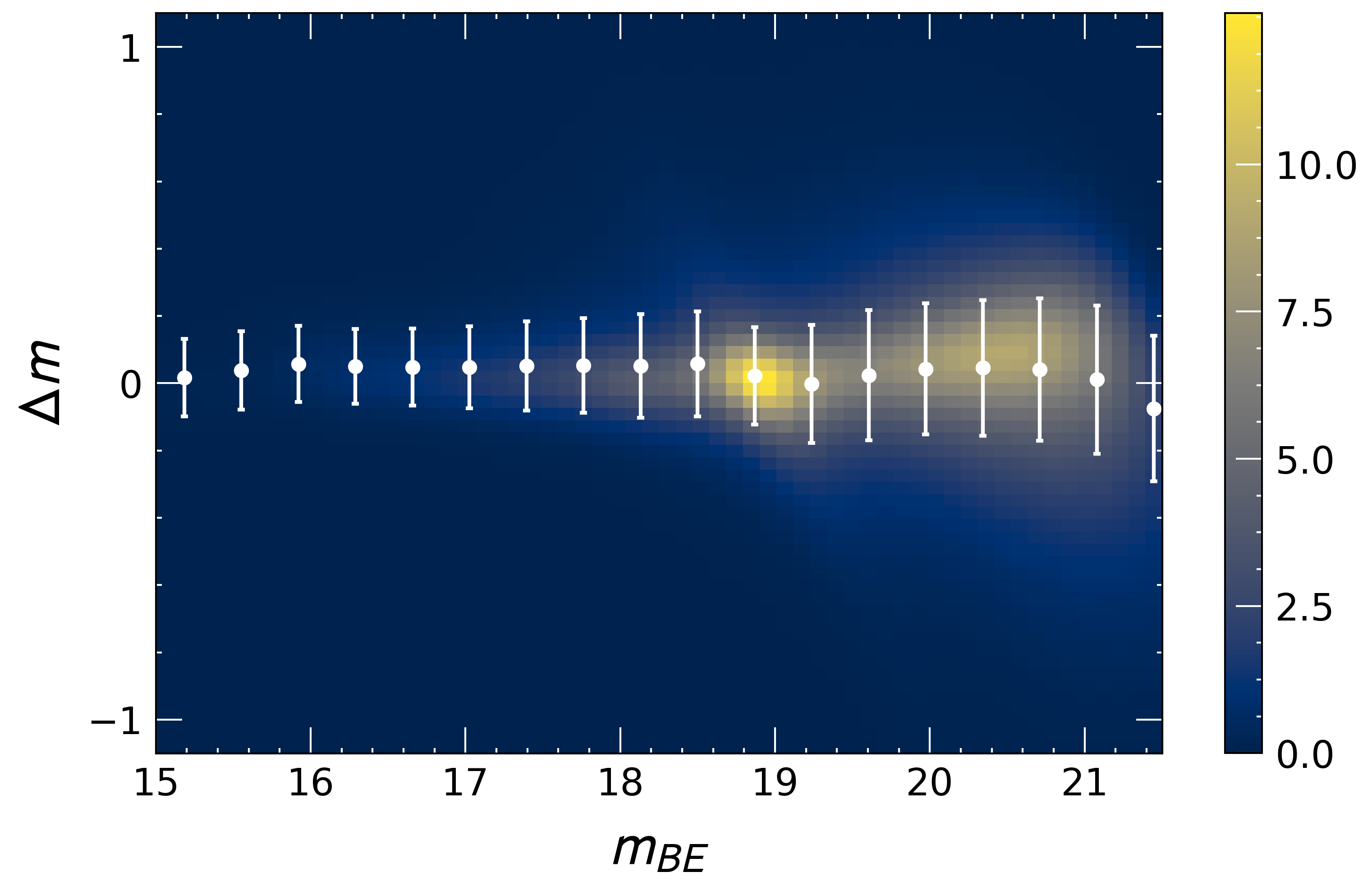}
    \caption{
    Distribution of the EROS-2 and MACHO magnitude difference $\Delta m=m_{BE}-m_{BM}$ as a function of the EROS-2 blue magnitude $m_{BE}$ for the 14.1 million of associated objects; the values of $m_{BM}$ are derived from the original MACHO magnitudes using a first-degree color equation, so as to match the EROS blue magnitudes on average.
    As a result, the mean value of $\Delta m$ varies by a few percent depending on the color of the sources, in particular between the main sequence and the red giants branch (around $m_{BE}=18.8$).
    White dots and bars show the average and standard deviations of $\Delta m$ for each $m_{BE}$ slice. Color scale is expressed in million of objects per squared magnitude.
    }
    \label{fig:assoc}
\end{figure}

To associate objects in the two catalogs, available MACHO and EROS-2 sky coordinates $(RA, DEC)$ were first refined using {\it Gaia} EDR3 astrometry \citep{2020Gaia_arXiv}, correcting for local shifts up to $2"$ for MACHO and $0.5"$ for EROS-2. 
After this correction, we could associate the objects of the two surveys with a precision of better than $0.1"$.
Given the typical spread of the light of the stars on the best images ($FWHM\sim 1"$), and the similar resolutions of the surveys, the associated reconstructed objects of each survey contain the same stars, with little variation of the blend components. 
This is confirmed by the good correlation observed between the EROS-2 and MACHO fluxes (Fig. \ref{fig:assoc})
which is compatible with the photometric accuracy.

Table \ref{Table-catalogue} summarizes the characteristics of the two surveys and the cross-matched catalog.
The common set consists of the $14.1\times10^6$ objects cross-matched in the EROS-2 and MACHO catalogs; 
they benefit from a total of 10.6 years of luminosity measurements, of which 3.8 years overlap, during which the surveys used 4 different passbands.
The complementary set, consisting of objects reconstructed in EROS-2 only or in MACHO only, comprising $22.7\times10^6$ objects monitored by a single survey, is also included in our analysis, although the objects are monitored in only 2 passbands for shorter times.

\begin{table}[h!]
\begin{center}
\begin{tabular}{|l|c c c |}\hline
    & EROS-2 only & MACHO only & common \\
    \hline
    Dates (month/yr) & 7/96-2/03 & 7/92-1/00 & 7/92-2/03 \\
    $T_{obs}$ (year) & 6.7 & 7.7 & 10.6 \\
    $N_{objects} (\times10^6)$ & 15.8 & $6.9^{(a)}$ & 14.1 \\
    \hline
    \footnotesize{\bf central fields deg$^2$} & $\sim 10$ & $\sim 10$ & $\sim 10$ \\
    $\rm stars/arcmin^2$ & $\sim 70$ & $\sim 100$ & $\sim 70$ \\
    mag. lim. $V_{Cousins}$ & $\sim 20.5$ & $\sim 20.5$ & $\sim 20.5$ \\
    \# measurements B & 500 & 1400 & 1900 \\
   \# measurements R & 600 & 1550 & 2150 \\
    \hline
    \footnotesize{\bf outer fields deg$^2$} & $\sim 77$ & $\sim 39$ & $\sim 39$ \\
    $\rm stars/arcmin^2$ & $\sim 30$ & $\sim 20$ & $\sim 20$ \\
    mag. lim. $V_{Cousins}$ & $\sim 22.5$ & $\sim 21.5$ & $\sim 21.5$ \\
    \# measurements B & 250 & 200 & 450 \\
   \# measurements R & 300 & 250 & 550 \\
   \hline
   \end{tabular}
   \centering
    \caption[]{Statistics of the EROS-2 and MACHO surveys.
    Survey durations, number of monitored sources, median stellar densities, approximate limiting magnitudes, median numbers of flux measurements per object after cleaning.
    $^{(a)}$ {\it number of unique objects}.
    }
    \label{Table-catalogue}
\end{center}
\end{table}

\subsection{Removing problematic images and measurements}
We found that bad images and/or measurements still polluted our light curves sample. MACHO images with more than $5\%$ of measurements more than $5$ standard deviations ($5\sigma$) away from the reference magnitude were found to be mostly faulty (visual inspection showed blurred images or with guiding or readout defect), and we then discarded them from our analysis.
We also rejected low quality EROS-2 images using similar criteria \citep{these_Blaineau}.

We further discarded measurements that deviated by more than $5\sigma$ from the median flux in a sliding window of 5 consecutive measurements, unless this occurred in more than $10\%$ of the cases (so as not to discard curves with many large and rapid variations). This operation would be penalizing when searching for events
of very short duration, but it has no impact on the efficiency for events lasting more
than a few months.

These measurement-quality cuts removed
about $\sim 3\%$ ($\sim 1\%$) of the measurements from the MACHO (EROS-2) data. 

\subsection{Corrections on flux uncertainties}
By comparing the mean photometric uncertainties with the point-to-point flux dispersions along the light curves,
it appeared that the MACHO uncertainties were in general underestimated by an average factor of $\sim0.73$ while those of EROS-2 were incorrectly estimated by a factor varying from $\sim0.75$ (for stars brighter than $I=19$) to $\sim1.45$ (for $I>21$).
In order to balance the weights of each survey in the calculations of the goodness-of-fit statistics $\chi^2$,
we renormalized the squared uncertainties for each light curve (one per passband) with
a quantity $X$ that characterizes the average fluctuations around a global trend,
estimated from the differences between the measurements and their nearest neighbors:
\begin{equation}
X = \frac{1}{N-2}\sum^{N-1}_{i=2}\left(\phi(t_i)-\phi^{int}_i \right)^2/\sigma_{int}^2
\end{equation}
Here, $\phi(t_i)$ is the flux measured at time $t_i$, $\phi^{int}_i$ is the flux interpolated from measurements $i-1$ and $i+1$ at $t_i$, and $\sigma_{int}$ is the uncertainty on $(\phi(t_i)-\phi^{int}_i)$ deduced from the cataloged uncertainties.
The sum is over points with a precision better than $0.55$~mag.
Renormalizing squared uncertainties  by dividing them by $X$
generally made the $\chi^2$ per DOF near unity 
for fits assuming slow flux variations over time. 
As the correction factors to be applied to photometric uncertainties vary with magnitudes,
this renormalization procedure is not effective for light curves with 
large variations, specially for low fluxes with large uncertainties ($>0.55$ mag) in EROS-2 data.

\section{Search for long timescale microlensing events}
\label{section:selection}
Thanks to the increase in computing power available compared to the 1990s and 2000s, we were able to conduct an analysis essentially based on the comparison of the fit of a microlensing effect with that of a constant light curve, without preselection criteria.
More precisely, for each object we performed a simultaneous PSPL microlensing fit to the available light curves, with one base flux line per passband and a set of microlensing parameters $(t_0,u_0,t_E)$ common to all passbands (fit without blending or parallax).
The selections described below were also applied to the simulated events, as described in section \ref{section:efficiency}, in order to determine the analysis efficiency.

The impact of the Earth's rotation (parallax) around the Sun on the detection efficiency of multi-year events was discussed in \citep{Blaineau2020}. The effect on a tolerant search algorithm such as the one we describe below was found to be negligible. Nevertheless,
simulated light curves included the parallax effect, even though for the search in the data (real and simulated) we only tried to fit a PSPL microlensing effect.
\subsection{Selection of candidates}
The light curves of the $36.8\times10^6$ cataloged objects underwent the following selection process to identify long-duration microlensing candidates. This process was tuned with the simulation described in Section \ref{section:efficiency}.
\begin{itemize}
\item
We eliminate objects with fewer than 200 total measurement points, all passbands included, and with fewer than 50 measurements in at least both EROS-2 or both MACHO passbands.
This selection leaves $36\times 10^6$ objects.
\item
We require that the light curves of the object simultaneously fit well a PSPL microlensing event, with a global $\chi^2$ (summed for all passbands) significantly smaller than that for constant curves.
Specifically, we require
\begin{equation}
\Delta\chi^2 = \frac{\chi^2_\text{const.}-\chi^2_{ML}}{\chi^2_{ML}/N_\text{dof}}\frac{1}{\sqrt{2N_\text{dof}}} > 80, \ {\rm and} \ \frac{\chi^2_{ML}}{N_\text{dof}} < f(\Delta\chi^2),
\label{deltachi2}
\end{equation}
where the function
\begin{equation}
f(\Delta\chi^2)=1.44 + 0.26\log_{10}\left(\frac{\Delta\chi^2}{80}\right)
+ 0.23\log_{10}^2\left(\frac{\Delta\chi^2}{80}\right) 
\end{equation}
is tuned to accept $92.5\%$ of the simulated light curves for each given interval of $\Delta\chi^2$,
and is in the range $1.4<f(\Delta\chi^2)<3$ for our entire data set.
These criteria accept $352$ light curves,
and between $28\%$ and $48\%$ of the events simulated within the observation duration with $u_0<1$ and $100<t_E<1000$ days (see Sect. \ref{section:efficiency}).
\item
We require that the global microlensing fit
does not result
in a significant degradation of $\chi^2_{ML}$ compared to $\chi^2_{const.}$ for any of the four passbands.
Specifically, we require $\Delta\chi^2>-0.1$, where $\Delta\chi^2$ is restricted to each light curve.
This requirement is satisfied by all the simulated events selected so far, and $226$ light curves remain at this stage.
All known events from the MACHO and EROS publications remain at this stage,
since the aforementioned criteria did not select preferentially long-duration events.
\item
We require that events have a maximum well within the observing period:
$t_{min} + 200\ days < t_0 < t_{max}$, where $t_{min}$ and $t_{max}$ are the start and end dates of the light curves.
This asymmetric requirement, which allows obtaining sufficient information about the rise phase, eliminates transients and some long-period variable stars that usually have a shorter rise time than the fall time, leaving 148 light curves.
\item
We require that $t_E$ be in the range $100\ days < t_E < (t_{max}-t_{min})/2$.
This criterion rejects most supernovae and short-lived fluctuations, 
and ensures that the fitted variations are sufficiently within the observing period.

$34$ light curves remain after these two last requirements. Since these criteria apply to the time parameters, they have an impact on the simulation that varies strongly with the mass of the lenses (from $17\%$ relative acceptance for $1\Msol$ to $62\%$ for $1000\Msol$).
\item 
We eliminate so-called “blue bumpers”
by making a stricter selection on events in their region of the color-magnitude diagram.
These stars are Be class stars located in the blue and bright zone of the Color-Magnitude Diagram (CMD). They sometimes present asymmetric bumps with a faster rise than fall, and their luminosity variations, probably related to the dynamics of the decretion disk \citep{2007Bestars}, can extend over years.
We reject these artifacts by requiring that the fitted value of $u_0$ be less than $0.9$ for objects in the following CMD domain:
$I_{Cousins} < 19$ and $(V_{Johnson} - I_{Cousins}) < 0.5$.

\end{itemize}
28 candidate events remain after this selection.

\subsection{Remaining candidates: rejection of known artefacts}
At this point, there are still some physical phenomena that cause changes in the objects' brightness, which can be mistaken for long-duration microlensing effects.
We eliminate three types of objects.
\begin{itemize}
\item
Objects outside the region of the CMD containing $99\%$ of the stars:
We reject these outliers
to discard in particular the rarest objects, likely to show variability.
$24$ objects remain after this CMD-based rejection, at the cost of $1\%$ of the expected events.
\item
Echoes from SN1987A \citep{SN1987A}: We discard the objects located in a zone defined
by $83.670 \degree < \alpha < 84.064 \degree$ and $-69.34 \degree < \delta < -69.20 \degree$.
After this exclusion, $5$ objects remain, while almost $100\%$ of the simulated events selected so far are retained.
\item
Variable objects identified in external catalogs:
At the end of our selection process, we reject one object at $(\alpha,\delta)_{J2000}=(80.6256,-71.7500)$, identified on external catalogs as a QSO \citep{Kim2012,Kozlowski2012}, one at $(76.7367,-71.4573)$ as a Young Stellar Object \citep{Whitney2008}, and one at $(87.9522,-74.5234)$ as a clear supernova, associated with a cataloged host-galaxy (LEDA database, \citet{Paturel1995}).
\end{itemize}

\begin{figure*}
    \centering
    \includegraphics[width=\textwidth]{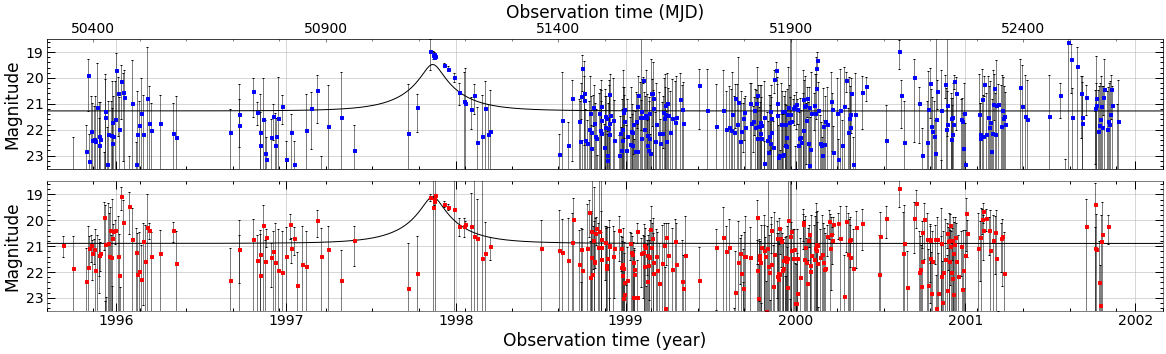}
    \caption{
    Light curves (magnitude vs. time) of candidate lm0690k17399 in EROS blue passband (upper panel), and EROS red passband (lower panel).
    The black solid lines show the best no-blend microlensing fit.
    No data is available in the MACHO catalog.
    The chromaticity, asymmetric shape, and large flux variation indicate that this is most likely a Type II-L or II-P supernova rather than a microlensing event.
    }
    \label{fig:LC-candidates}
\end{figure*}

\begin{figure*}
    \centering
    \includegraphics[width=\textwidth]{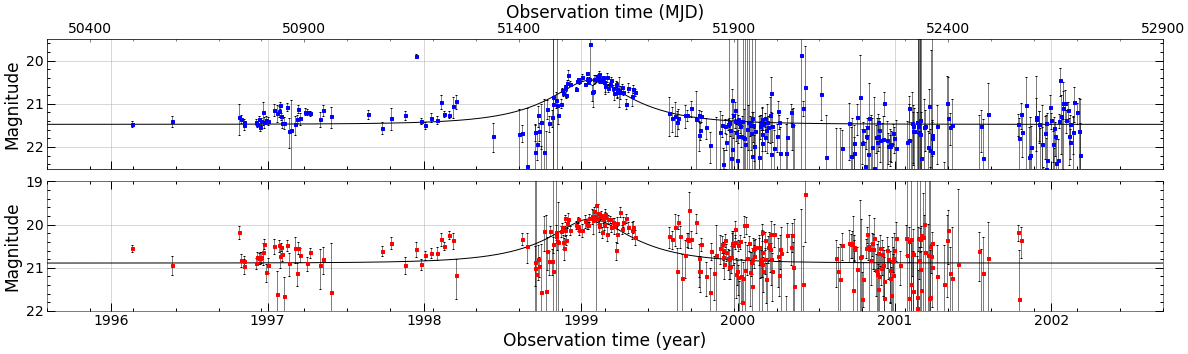}
    \caption{
    Light curves (magnitude vs. time) of candidate lm0073m17729 in EROS blue passband (upper panel), and EROS red passband (lower panel).
    The black solid lines show the best no-blend microlensing fit.
    No data is available in the MACHO catalog.
    The hints of variability outside the main bump and the asymmetry, especially visible in the change of the blue baseline after the maximum, make it unlikely that this event is a geniune microlensing event.
    }
    \label{fig:LC-candidate_2}
\end{figure*}

After elimination of these objects, only two candidates remain.
Both were observed by EROS-2 but are not within  any MACHO field.
Their characteristics and light curves are shown in Table \ref{tab:candidates} and Figs. \ref{fig:LC-candidates} and \ref{fig:LC-candidate_2}.

Candidate {\it lm0690k17399} is one of the $2\%$ faintest objects in the EROS-2 catalog, and
$\approx20\%$ of its measurements had negative flux.
The event itself  is chromatic and has an asymmetric shape with a factor of flux increase $>5$ (probably even larger when blend is taken into account),
characteristic of a  type II-L or II-P supernovae \citep{SNIIL}.
Differential photometry indicates that the object's position during
the event was offset from the position outside the event by $\approx0.6"$, again being consistent with being a supernova in a background galaxy.

Candidate {\it lm0073m17729} shows hints of variability outside the main event, and the post-event baseline in the blue appears to be $\approx0.4$~mag fainter than the pre-event baseline.

The light curves of these candidates bear little resemblance to the light curves of actual microlensing effects.
However, in our analysis, we cannot formally exclude these candidates without additional data or stricter selection.
Therefore, we have chosen to consider them, but only to establish upper bounds, and not to derive a microlensing optical depth due to halo compact objects.
We will show in Sect. \ref{fig:resultat} that considering or not these two candidates has no consequence on the upper limit of the contribution of high mass compact objects (heavier than $20\Msol$) to the halo.
\begin{table}
    \begin{center}
\begin{tabular}{ccc} \hline \\  [-1ex]
candidate ID & lm0690k17399 & lm0073m17729 \\
\hline
RA (J2000) & 78.3674 & 93.3520 \\
DEC (J2000) & -71.9644 & -69.5183  \\
R & $20.89\pm0.04$ & $20.89\pm0.02$\\
B & $21.28\pm0.06$ & $21.48\pm0.01$ \\
$u_0$ & $0.194\pm0.010$ & $0.413\pm0.007$\\
\vspace{1mm}
$t_E$ (day) & $106.3^{+9.6}_{-8.0}$ & $183.1^{+8.3}_{-7.8}$ \\
$t_0$ (MJD) & $51129.7^{+1.6}_{-1.9}$ & $51567.2\pm2.0$ \\
\vspace{1mm}
$\Delta\chi^2$ & 106.16 & 85.02 \\
 \hline
\end{tabular}
\centering
\caption[]{Characteristics of the candidates, with fitted microlensing parameters.
}
\label{tab:candidates}
\end{center}
\end{table}


\section{Efficiency and expected detection rate}
\label{section:efficiency}
For $N_{objects}$ objects observed over a time , $T_{obs}$, 
the expected number of detected events is
\begin{equation}
N_{exp}=N_{objects} T_{obs}\int dt_E \,\epsilon(t_E) \,\frac{d\Gamma}{dt_E}
\label{eqn:nexpected}
\end{equation}
where $d\Gamma/dt_E$ is the rate per $t_E$ interval and $\epsilon(t_E)$
is the detection efficiency, corresponding to the fraction of microlensing events accepted by the selection cuts, relative to the events for which $t_0$ is within the observation time $T_{obs}$ and $u_0<1$.
Because of the “blending” effects, 
there is a non-trivial relation between the number of objects and the
number of stars, so Eq. (\ref{eqn:nexpected})
effectively defines $\epsilon(t_E)$.
As explained in the following subsections,
the number $N_{exp}$ depends on the details of the observations,
on the blending of stars, and on the halo model.

\subsection{Simulations and efficiency}

The efficiency, $\epsilon(t_E)$ was determined by superimposing microlensing
events on the light curves of a representative random subsample of the observed objects and then subjecting
these new simulated light curves to the standard analysis procedure.
Two types of simulations were performed. The first assumed that each observed
object corresponds to one and only one star and that variations of the
associated light curves can be induced only by microlensing of that one star.
The second, more realistic simulations, takes into account “blending”
by assigning a set of stars to each object in a way that is consistent
with the observed density of LMC stars in HST images, as described in the next subsection.

We generate microlensing events by modifying the light curves of a 
random sub-sample of  objects as follows.
First, for each object values of $t_E$, $u_0$, $t_0$ and parallax parameters
are drawn randomly from the appropriate distributions. 
In particular,
$u_0$ is drawn uniformly in the interval $0<u_0<1.5$.
The time of event maximum, $t_0$, is generated randomly over
a time period that extends sufficiently beyond the first and last observations, 
since photometric fluctuations may make such events appear to be within the
observing period.
The $t_E$ distribution is derived from the mass, distance and velocity distributions of the lenses.
The parallax depends in addition
on the random orientation of the transverse velocity vector of the lens.
For each of the $N_{star}$ stars associated with the object, 
modified object light curves
are constructed by modifying the star's flux according to the theoretical
microlensing light curve and then modifying the object light curves
in the appropriate manner, taking into account the star's contribution to the object.
The uncertainties associated with the flux measurements in the object light curves are modified to account for the increased flux.
Each of the $N_{star}$ set of light curves (or event) is then subjected to the selection process described in the previous section.
The efficiency $\epsilon(t_E,u_0)$ is then the number of simulated events passing the cuts divided by the number of objects.
Because in the simulation with blending there is more than one star per object,
$\epsilon(t_E,u_0)$ can be greater than unity, essentially near $u_0=0$.

The calculation of the expected number of events requires $\epsilon(t_E)$ 
defined as the number of simulated events passing the cuts divided by the number of objects whose event was simulated with $u_0<1$.
Figure \ref{fig:tEdist_eff} shows the resulting efficiency on simulations with and without blending for the analysis described in
Sect. \ref{section:selection}, and also by imposing the additional constraint $t_E>200$ days.
The effects of blending are similar to the modest effects seen in earlier
studies.
Blending increases the true number of stars subject to lensing, but
it also affects the shape of the light curve, by
decreasing the apparent duration $t_E$ and increasing the apparent impact parameter $u_0$.
The decrease in effective $t_E$ lowers the efficiency for events at low $t_E$
but increases it at high $t_E$.
This change in efficiency is seen in Fig. \ref{fig:tEdist_eff}.

\subsection{Blending}
The impact of blending was quantified in the simulation as a multiplicative factor of the number of amplifiable objects.
We statistically assign to each object a list of HST contributing stars with their positions, and calculate the contribution of each star similarly to \citep{Tisserand_2007} and \citep{Wyrzykowski_2011}, which relied on artificial star additions.
In dense fields near the LMC bar, we find that on average 2.07 HST stars contribute at least $10\%$ of the flux of cataloged objects (1.58 for objects brighter than I=19.5). In the sparsest fields, this number drops to 1.46 (1.1 for bright objects).
Overall, we find that 90\% of the events passing all cuts are due to microlensing of the brightest star associated with an object.

This procedure, based on the HST image analysis, is sufficient
as long as the spatial distribution of the blend constituents is random, as appears to be the case in the LMC HST images;
but a statistical analysis of binary systems closer than $600\,$pc extracted from the {\it Gaia} database \citep{these_Blaineau} showed us that such systems, located at the distance of the LMC ($49.5\,$kpc), increase the chances that a star has a neighbor not resolved by HST.
We therefore decided to examine further this particular regime of blend.
In \citet{these_Blaineau} and in a forthcoming paper, we show that the components of
most of the LMC binaries are too close together for them to experience very different magnifications through a heavy lens (which almost always have $r_E>50AU$ when $M_L>100\Msol$).
We indeed find that less than $7\%$ of the LMC objects should be binaries unresolved by the HST with a projected orbital distance $>50AU$.
Because of this, we ignore the effect in the calculation of the efficiency.

Figure \ref{fig:tEdist_eff} shows the detection efficiency for the analysis just described (labelled $t_E>100$ days) as well as the efficiency for the same analysis, but with a stricter requirement on the event duration  $t_E>200$ days.
\subsection{The expected number of detected events}
\begin{figure}
    \centering
    \includegraphics[width=\linewidth]{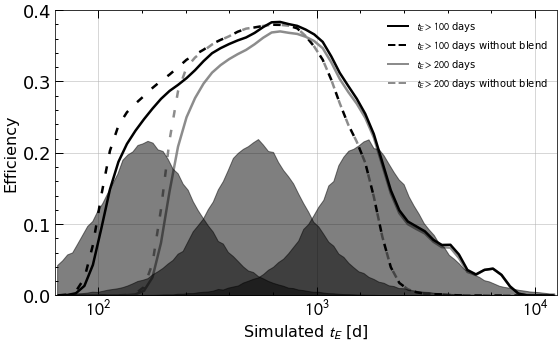}
    \caption{
    Detection efficiency:
     Solid (dashed) lines show efficiencies, taking (or not) into account blending effects. 
    Grey lines show efficiencies of the analysis adding the constraint $t_E>200$ days.
Grey histograms (not normalized) are the expected $t_E$ distributions for a halo made of $10$, $100$, and $1000\Msol$ compact objects (from left to right).
    }
    \label{fig:tEdist_eff}
\end{figure}

Top panel of figure \ref{fig:resultat} shows the number of detected microlensing
events $N_{exp}^{(h)}(M)$ expected from lenses in the standard isothermal spherical halo (S-model) with $\tau_{LMC}\sim4.7\times 10^{-7}$, assumed to all have the same mass $M$. 
This number is calculated as a function of $M$ from Eq. (\ref{eqn:nexpected}), using the efficiency shown in Fig. \ref{fig:tEdist_eff} for the simulation with blending.
For the number of source objects $N_{objects}$,
we subtract from the $36.8\times 10^6$ objects in our catalog the contamination by Galactic stars, which we estimated to be less than $5\%$ by counting stars in the {\it Gaia} catalog of fields located at the same Galactic latitude, but away from the LMC \citep{2020Gaia_arXiv}.
Finally, as in \citet{Tisserand_2007} and \citet{Wyrzykowski_2011}, we consider that $10\%$ of microlensing events may escape detection due to lens binarity \citep{Mroz2019}, to conservatively infer the expected number of events $N_{exp}^{(h)}(M)$ as a function of the deflector's mass $M$.
For a wide  mass distribution, the number of expected events from halo lenses can simply be calculated 
by integrating the $N_{exp}^{(h)}(M)$ curve, weighted by the deflector mass distribution function.

The expected number of detected events (Eq. \ref{eqn:nexpected})
varies with the adopted halo model
through the optical depth, $\tau_{LMC}$, and $t_E$ distributions.
The range of plausible models was discussed in 
\citep{1996ApJ...461...84A} and we refer to their Table 2 for model details.
Isothermal models consistent with flat rotation curve, e.g. 
model S of \citet{1996ApJ...461...84A},
all yield values of  $N_{exp}^{(h)}$ similar to that of our baseline model, 
while $N_{exp}^{(h)}$ is significantly reduced for
models with a falling rotation curve, as might be suggested by some Milky Way rotation 
velocity measurements (e.g. \citet{2014ApJ...785...63B}).
For the model C of \citet{1996ApJ...461...84A} with $\tau_{LMC}\approx3.0\times10^{-7}$, and $\langle t_E\rangle\approx 120\,day \sqrt{M/M_\odot}$, $N_{exp}^{(h)}$
is decreased  by a factor $\approx 2$ 
for $M\approx10 M_\odot$  and the maximum mass for which one would 
expect three events is reduced by a factor $\approx 3$.

\section{New limits on compact objects in the halo}
\label{section:resultats}
We have taken into account the contribution of the Galactic disk lenses and the LMC self-lensing by adapting the results of \citet{CalchiNovati2011} to our efficiency estimate. We therefore expect $N_{exp}^{(s)}\sim 0.64$ disk and self-lensing events with $t_E>100$ days to be compared with 2 observed candidates.
In the same way as \citep{CalchiNovati2011} we perform Bayesian inference, assuming a weakly constraining uniform prior for the optical depth of the halo compact objects 
$0 \leq \tau_{LMC} \leq 4.7 \times 10^{-6}$ 
({\it i.e.} up to 10 times the expected maximum optical depth).
For $N_{obs}$ observed events with durations $t_i  \, , i=1 \ldots N_{obs}$, the likelihood for the fractional optical depth $f=\tau_{LMC}/4.7 \times 10^{-7}$ is defined for a given lens mass $M$ as: 
\begin{eqnarray}
    \mathcal{L}(f ; M | obs ) =
\frac{e^{-N_{exp}} }{N_{obs} !}    \prod_{i=1}^{N_{obs}}  \left[\frac{d\Gamma_{exp}^{(s)}}{dt_E} (t_i)\! +\! f\! \times \! \frac{d\Gamma_{exp}^{(h)} (M)}{dt_E} (t_i) \right] \\
    \int d\Gamma_{exp}^{(s)} = N_{exp}^{(s)} \hspace{6mm} \int d\Gamma_{exp}^{(h)} = N_{exp}^{(h)} \nonumber \\
    N_{exp}  =  N_{exp}^{(s)} + f \times N_{exp}^{(h)} \nonumber 
\end{eqnarray}
where $d\Gamma_{exp}^{(s)}/dt_E$ and $d\Gamma_{exp}^{(h)}(M)/dt_E$ are the differential microlensing event rates due to the Milky Way disk and the LMC self-lensing ($\Gamma_{exp}^{(s)}$), and due to a halo  ($\Gamma_{exp}^{(h)}$) completely made of compact objects with identical mass $M$; $N_{exp}^{(s)}$ and $N_{exp}^{(h)}$ are 
the corresponding numbers of expected events.
The probability of observing the two events (with $t_E\,=106\,$days and $t_E\,=183\,$days), for a given halo fraction $f$ is then:
\begin{equation}
p(obs | f) = \frac{ \mathcal{L}(f; M | obs ) }{ \int_0^{10} \mathcal{L}(f; M | obs ) \, d f }
\end{equation}

\begin{figure}
    \centering
    \includegraphics[width=0.9\linewidth]{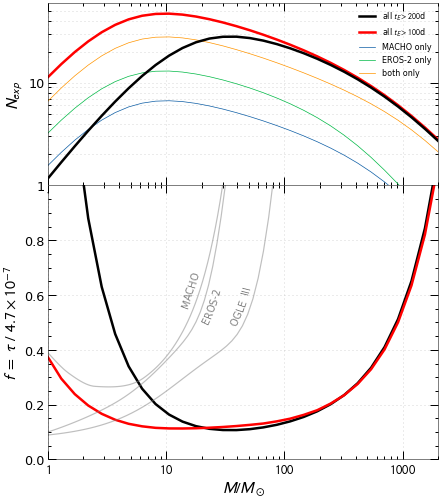}
    \caption{
Top:
Number of events expected from a halo S-model entirely composed of compact objects of mass $M$:
blue (green) line, from source objects monitored only by MACHO (EROS-2);
orange line, from source objects monitored by both surveys; full red line shows the total; black line shows the total adding the constraint $t_E>200$ days in the analysis.
Bottom: $95\%$ CL upper limits on the fraction of the halo mass in the form of compact objects $f=\tau_{LMC}/4.7\times 10^{-7}$.
Limits obtained in this analysis are shown 
in red, and in black if we require $t_E>200$ days.  The gray curves correspond to the latest limits published by MACHO, EROS-2 and OGLE-III.
    }
    \label{fig:resultat}
\end{figure}

The halo fraction excluded with 95\% CL through this Bayesian analysis,
as a function of deflector mass $f(M)$, is shown on the bottom panel of Fig. \ref{fig:resultat} as the red curve.  
Assuming $N_{exp}^{(s)}=0$ has a negligible impact on the exclusion curve, as the expected 
signal from the halo is dominant, as shown in the upper part of the figure. 

We tested the robustness of this result by changing the $t_E>100\,$days requirement
to the stricter $t_E>200\,$days cut. In doing so $N_{exp}^{(s)}$ becomes negligible ($<0.05$ event) \citep{CalchiNovati2011}, and we are left with zero event in the data. 
The resulting exclusion limit, using Poisson statistics, shown as the black curve,
is weaker at the lower mass end, but unchanged on the high mass side. 
We have also shown as gray curves the previously published excluded halo fractions
by MACHO \citep{MACHO_2001}, EROS-2 \citep{Tisserand_2007},
and OGLE-III \citep{Wyrzykowski_2011}.

To test halos with mass distribution extending below $1M_\odot$, we can use
the results of EROS-1 \citep{1997A&A...324L..69R} and EROS-2 reported in \citet{Tisserand_2007}.
They expected $\gtrsim 40$ events over a wide range $10^{-6}<M<1M_\odot$
peaking at 120 events for $10^{-2}M_\odot$, whereas no events were seen.
Combining the results of the analysis reported here with those of \citep{Tisserand_2007}
we would expect $\approx40$ events for a flat distribution of $\log M$ extending
over the range $10^{-7}<M<10^3M_\odot$. 
More generally, the combination of the limits obtained by EROS-1, EROS-2, and the present analysis shows that deflectors with any mass distribution in the range $10^{-6}<M<10^2M_\odot$ 
can not contribute more than $f \sim 15\%$ at 95\% CL to the halo mass, assuming a standard 
spherical halo.

Constraints on the existence of primordial black holes
are reviewed in \citet{2021RPPh...84k6902C}. 
Limits at the level of 10\% on the halo fraction have been established for
most of the range $10^{-10}<M<10^2M_\odot$.
At higher masses, $10^3<M<10^{15}M_\odot$, a variety of techniques have provided
stronger limits, as low of $10^{-4}$.
At lower masses, primordial Black Holes with $M<10^{-17}M_\odot$
are ruled out because of their rapid evaporation via Hawking radiation.

\section{Discussion, conclusions, and perspectives}
\label{section:discussion}
In this search, we did not consider the SMC data because of their low statistical impact (5.2 million light curves).

We attribute the significant improvement in sensitivity over previously reported results to several factors:
\begin{itemize}
    \item 
    Past analysis focused on finding shorter duration events, and always imposed strong constraints on the maximum value of $t_E$
    to allow checking the stability of the object outside the microlensing effect.
    \item
    The addition of the expected signals in EROS-2 and MACHO, taking care to count only once the objects followed simultaneously, already mechanically improves the sensitivity \citep{Moniez2010}.
    Note also that we were able to use more archived data (7.7 years) from MACHO than the data analyzed by this collaboration (5.7 years).
    \item
    A significant part of the improvement comes from combining the light curves, extended to 10.6 years, for the 14.1 million objects jointly observed by EROS-2 and MACHO.
    These objects, which represent $38\%$ of the catalog, contribute for $4.0/5.55=72\%$ of the expected events if all the lenses have a mass of $1000\Msol$ (Fig. \ref{fig:resultat}-Top). This indicates that the detection efficiency on the jointly measured curves is increased by a factor of $\sim 4$ compared to objects measured by only one survey.
\end{itemize}
The combined data set from EROS-2, MACHO and OGLE surveys
now spans 30 consecutive years,
since OGLE-III observed from sept. 2001 to march 2009 and OGLE-IV started in march 2010. It therefore still has potential, which has been studied in \citet{Mirhosseini2018}. A group that would analyze all these surveys on the search principle applied here of simultaneous fitting of all light curves in all available passbands (up to 6) could certainly further improve the limits obtained here, or result in the detection of some very long duration events.
\citet{Mirhosseini2018} predicted that if the halo was completely made up of $1000\Msol$ objects there should be at least 3 microlensing events towards LMC in the combined data from all previous microlensing search programs (EROS-2, MACHO, OGLE through 2018).
In fact, these predictions were based on a conservative extrapolation of the detection efficiency, and the combination of only EROS-2 and MACHO data, as analysed here, allows us to conclude that $1000\Msol$ objects filling the halo should have produced $\sim 6$ detectable events.
The limit of this catalog fusion technique will be reached when all existing catalogs (EROS-2+MACHO+OGLE completed) have been used.
If an analysis of a combined catalog with 20 millions objects monitored for 30 years has a mean microlensing detection efficiency of 0.20, and assuming no event is detected, then the (ultimate) limit on the mass of the compact halo objects as unique component of the standard halo could be pushed to $\gtrsim 3000 \Msol$.
The Rubin/LSST survey \citep{2019ApJ...873..111I}, which will further extend the monitoring duration of the LMC objects with better photometric precision (also allowing for better detection efficiency), should further enhance the value of the historical surveys whose data could also be aggregated, or at least used for verification purposes in case of detection.

In conclusion, it seems that BHs with masses up to a thousand solar masses, similar to the ones observed by LIGO and Virgo as binary BH mergers, do not make up a major fraction of the Milky Way dark matter, at least if assumed to be distributed as a standard spherical halo. Such BHs are more likely to be found in structures following the visible mass distribution, and could be searched for through microlensing toward the Galactic bulge and spiral arms, in the long $t_E$ tail of event duration distribution,
extending the previous searches and analysis \citep{Hamadache2006, Moniez2017, Mroz2020}.

\begin{acknowledgements}
We thank
Lucasz Wyrzykowski and Christopher Stubbs for providing useful information. This work was supported by the Paris Ile-de-France Region.\\
This paper uses public domain data obtained by the MACHO Project, jointly funded by the US Department of Energy through the University of California, Lawrence Livermore National Laboratory under contract No. W-7405-Eng-48, by the National Science Foundation through the Center for Particle Astrophysics of the University of California under cooperative agreement AST-8809616, and by the Mount Stromlo and Siding Spring Observatory, part of the Australian National University.\\
This article is dedicated to the memory of our EROS collaborators 
Johannes Andersen,
Pierre Bareyre,
Florian Bauer,
Sergio Char,
Eric Maurice,
Alain Milsztajn,
Luciano Moscoso, and
C\'ecile Renault.

\end{acknowledgements}
\bibliographystyle{aa}
\bibliography{citations_microlensing-combined}

\begin{thebibliography}{39}
\expandafter\ifx\csname natexlab\endcsname\relax\def\natexlab#1{#1}\fi

\bibitem[{Abbott {et~al.}(2016{\natexlab{a}})Abbott, Abbott, Abbott, Abernathy,
  Acernese, Ackley, Adams, Adams, Addesso, Adhikari, \& et~al.}]{Abbott_2016b}
Abbott, B., Abbott, R., Abbott, T., {et~al.} 2016{\natexlab{a}}, \prl, 116,
  241103

\bibitem[{Abbott {et~al.}(2016{\natexlab{b}})Abbott, Abbott, Abbott, Abernathy,
  Acernese, Ackley, Adams, Adams, Addesso, Adhikari, \& et~al.}]{Abbott_2016a}
Abbott, B., Abbott, R., Abbott, T., {et~al.} 2016{\natexlab{b}}, \prl, 116,
  061102

\bibitem[{{Alcock} {et~al.}(2001){Alcock}, {Allsman}, {Alves}, {Axelrod},
  {Becker}, {Bennett}, {Cook}, {Dalal}, {Drake}, {Freeman}, {Geha}, {Griest},
  {Lehner}, {Marshall}, {Minniti}, {Nelson}, {Peterson}, {Popowski}, {Pratt},
  {Quinn}, {Stubbs}, {Sutherland}, {Tomaney}, {Vandehei}, \&
  {Welch}}]{MACHO_2001}
{Alcock}, C., {Allsman}, R.~A., {Alves}, D.~R., {et~al.} 2001, \apjl, 550, L169

\bibitem[{{Alcock} {et~al.}(1999){Alcock}, {Allsman}, {Alves}, {Axelrod},
  {Becker}, {Bennett}, {Cook}, {Drake}, {Freeman}, {Geha}, {Griest}, {Lehner},
  {Marshall}, {Minniti}, {Peterson}, {Popowski}, {Pratt}, {Nelson}, {Quinn},
  {Stubbs}, {Sutherland}, {Tomaney}, {Vandehei}, {Welch}, \& {MACHO
  Collaboration}}]{Alcock1999}
{Alcock}, C., {Allsman}, R.~A., {Alves}, D.~R., {et~al.} 1999, \pasp, 111, 1539

\bibitem[{{Alcock} {et~al.}(1996){Alcock}, {Allsman}, {Axelrod}, {Bennett},
  {Cook}, {Freeman}, {Griest}, {Guern}, {Lehner}, {Marshall}, {Park},
  {Perlmutter}, {Peterson}, {Pratt}, {Quinn}, {Rodgers}, {Stubbs}, \&
  {Sutherland}}]{1996ApJ...461...84A}
{Alcock}, C., {Allsman}, R.~A., {Axelrod}, T.~S., {et~al.} 1996, \apj, 461, 84

\bibitem[{{Bhattacharjee} {et~al.}(2014){Bhattacharjee}, {Chaudhury}, \&
  {Kundu}}]{2014ApJ...785...63B}
{Bhattacharjee}, P., {Chaudhury}, S., \& {Kundu}, S. 2014, \apj, 785, 63

\bibitem[{{Bird} {et~al.}(2016){Bird}, {Cholis}, {Mu{\~n}oz},
  {Ali-Ha{\"\i}moud}, {Kamionkowski}, {Kovetz}, {Raccanelli}, \&
  {Riess}}]{Bird_2016}
{Bird}, S., {Cholis}, I., {Mu{\~n}oz}, J.~B., {et~al.} 2016, \prl, 116, 201301

\bibitem[{Blaineau(2021)}]{these_Blaineau}
Blaineau, T. 2021, PhD thesis

\bibitem[{{Blaineau} \& {Moniez}(2020)}]{Blaineau2020}
{Blaineau}, T. \& {Moniez}, M. 2020, A\&A, 636, L9

\bibitem[{{Brunthaler} {et~al.}(2011){Brunthaler}, {Reid}, {Menten}, {Zheng},
  {Bartkiewicz}, {Choi}, {Dame}, {Hachisuka}, {Immer}, {Moellenbrock},
  {Moscadelli}, {Rygl}, {Sanna}, {Sato}, {Wu}, {Xu}, \&
  {Zhang}}]{Brunthaler_2011}
{Brunthaler}, A., {Reid}, M.~J., {Menten}, K.~M., {et~al.} 2011, Astronomische
  Nachrichten, 332, 461

\bibitem[{{Calchi Novati} \& {Mancini}(2011)}]{CalchiNovati2011}
{Calchi Novati}, S. \& {Mancini}, L. 2011, \mnras, 416, 1292

\bibitem[{{Carr} {et~al.}(2021){Carr}, {Kohri}, {Sendouda}, \&
  {Yokoyama}}]{2021RPPh...84k6902C}
{Carr}, B., {Kohri}, K., {Sendouda}, Y., \& {Yokoyama}, J. 2021, Reports on
  Progress in Physics, 84, 116902

\bibitem[{{Gaia Collaboration} {et~al.}(2020){Gaia Collaboration}, {Brown},
  {Vallenari}, {Prusti}, {de Bruijne}, {Babusiaux}, \&
  {Biermann}}]{2020Gaia_arXiv}
{Gaia Collaboration}, {Brown}, A.~G.~A., {Vallenari}, A., {et~al.} 2020, arXiv
  e-prints, arXiv:2012.01533

\bibitem[{{Green} \& {Kavanagh}(2021)}]{2021JPhG...48d3001G}
{Green}, A.~M. \& {Kavanagh}, B.~J. 2021, Journal of Physics G Nuclear Physics,
  48, 043001

\bibitem[{{Griest}(1991)}]{1991ApJ...366..412G}
{Griest}, K. 1991, \apj, 366, 412

\bibitem[{{Hamadache} {et~al.}(2006){Hamadache}, {Le Guillou}, {Tisserand},
  {Afonso}, {Albert}, {Andersen}, {Ansari}, {Aubourg}, {Bareyre}, {Beaulieu},
  {Charlot}, {Coutures}, {Ferlet}, {Fouqu{\'e}}, {Glicenstein}, {Goldman},
  {Gould}, {Graff}, {Gros}, {Haissinski}, {de Kat}, {Lesquoy}, {Loup},
  {Magneville}, {Marquette}, {Maurice}, {Maury}, {Milsztajn}, {Moniez},
  {Palanque-Delabrouille}, {Perdereau}, {Rahal}, {Rich}, {Spiro},
  {Vidal-Madjar}, {Vigroux}, \& {Zylberajch}}]{Hamadache2006}
{Hamadache}, C., {Le Guillou}, L., {Tisserand}, P., {et~al.} 2006, \aap, 454,
  185

\bibitem[{{Hubert}(2007)}]{2007Bestars}
{Hubert}, A.~M. 2007, in Astronomical Society of the Pacific Conference Series,
  Vol. 361, Active OB-Stars: Laboratories for Stellare and Circumstellar
  Physics, ed. A.~T. {Okazaki}, S.~P. {Owocki}, \& S.~{Stefl}, 27

\bibitem[{{Ivezi{\'c}} {et~al.}(2019){Ivezi{\'c}}, {Kahn}, {Tyson}, {Abel},
  {Acosta}, {Allsman}, {Alonso}, {AlSayyad}, {Anderson}, {Andrew}, {Angel},
  {Angeli}, {Ansari}, {Antilogus}, {Araujo}, {Armstrong}, {Arndt}, {Astier},
  {Aubourg}, {Auza}, {Axelrod}, {Bard}, {Barr}, {Barrau}, {Bartlett}, {Bauer},
  {Bauman}, {Baumont}, {Bechtol}, {Bechtol}, {Becker}, {Becla}, {Beldica},
  {Bellavia}, {Bianco}, {Biswas}, {Blanc}, {Blazek}, {Blandford}, {Bloom},
  {Bogart}, {Bond}, {Booth}, {Borgland}, {Borne}, {Bosch}, {Boutigny},
  {Brackett}, {Bradshaw}, {Brandt}, {Brown}, {Bullock}, {Burchat}, {Burke},
  {Cagnoli}, {Calabrese}, {Callahan}, {Callen}, {Carlin}, {Carlson},
  {Chandrasekharan}, {Charles-Emerson}, {Chesley}, {Cheu}, {Chiang}, {Chiang},
  {Chirino}, {Chow}, {Ciardi}, {Claver}, {Cohen-Tanugi}, {Cockrum}, {Coles},
  {Connolly}, {Cook}, {Cooray}, {Covey}, {Cribbs}, {Cui}, {Cutri}, {Daly},
  {Daniel}, {Daruich}, {Daubard}, {Daues}, {Dawson}, {Delgado}, {Dellapenna},
  {de Peyster}, {de Val-Borro}, {Digel}, {Doherty}, {Dubois},
  {Dubois-Felsmann}, {Durech}, {Economou}, {Eifler}, {Eracleous}, {Emmons},
  {Fausti Neto}, {Ferguson}, {Figueroa}, {Fisher-Levine}, {Focke}, {Foss},
  {Frank}, {Freemon}, {Gangler}, {Gawiser}, {Geary}, {Gee}, {Geha}, {Gessner},
  {Gibson}, {Gilmore}, {Glanzman}, {Glick}, {Goldina}, {Goldstein}, {Goodenow},
  {Graham}, {Gressler}, {Gris}, {Guy}, {Guyonnet}, {Haller}, {Harris},
  {Hascall}, {Haupt}, {Hernandez}, {Herrmann}, {Hileman}, {Hoblitt}, {Hodgson},
  {Hogan}, {Howard}, {Huang}, {Huffer}, {Ingraham}, {Innes}, {Jacoby}, {Jain},
  {Jammes}, {Jee}, {Jenness}, {Jernigan}, {Jevremovi{\'c}}, {Johns}, {Johnson},
  {Johnson}, {Jones}, {Juramy-Gilles}, {Juri{\'c}}, {Kalirai}, {Kallivayalil},
  {Kalmbach}, {Kantor}, {Karst}, {Kasliwal}, {Kelly}, {Kessler}, {Kinnison},
  {Kirkby}, {Knox}, {Kotov}, {Krabbendam}, {Krughoff}, {Kub{\'a}nek},
  {Kuczewski}, {Kulkarni}, {Ku}, {Kurita}, {Lage}, {Lambert}, {Lange},
  {Langton}, {Le Guillou}, {Levine}, {Liang}, {Lim}, {Lintott}, {Long},
  {Lopez}, {Lotz}, {Lupton}, {Lust}, {MacArthur}, {Mahabal}, {Mandelbaum},
  {Markiewicz}, {Marsh}, {Marshall}, {Marshall}, {May}, {McKercher}, {McQueen},
  {Meyers}, {Migliore}, {Miller}, {Mills}, {Miraval}, {Moeyens}, {Moolekamp},
  {Monet}, {Moniez}, {Monkewitz}, {Montgomery}, {Morrison}, {Mueller},
  {Muller}, {Mu{\~n}oz Arancibia}, {Neill}, {Newbry}, {Nief}, {Nomerotski},
  {Nordby}, {O'Connor}, {Oliver}, {Olivier}, {Olsen}, {O'Mullane}, {Ortiz},
  {Osier}, {Owen}, {Pain}, {Palecek}, {Parejko}, {Parsons}, {Pease},
  {Peterson}, {Peterson}, {Petravick}, {Libby Petrick}, {Petry},
  {Pierfederici}, {Pietrowicz}, {Pike}, {Pinto}, {Plante}, {Plate}, {Plutchak},
  {Price}, {Prouza}, {Radeka}, {Rajagopal}, {Rasmussen}, {Regnault}, {Reil},
  {Reiss}, {Reuter}, {Ridgway}, {Riot}, {Ritz}, {Robinson}, {Roby}, {Roodman},
  {Rosing}, {Roucelle}, {Rumore}, {Russo}, {Saha}, {Sassolas}, {Schalk},
  {Schellart}, {Schindler}, {Schmidt}, {Schneider}, {Schneider}, {Schoening},
  {Schumacher}, {Schwamb}, {Sebag}, {Selvy}, {Sembroski}, {Seppala}, {Serio},
  {Serrano}, {Shaw}, {Shipsey}, {Sick}, {Silvestri}, {Slater}, {Smith},
  {Smith}, {Sobhani}, {Soldahl}, {Storrie-Lombardi}, {Stover}, {Strauss},
  {Street}, {Stubbs}, {Sullivan}, {Sweeney}, {Swinbank}, {Szalay}, {Takacs},
  {Tether}, {Thaler}, {Thayer}, {Thomas}, {Thornton}, {Thukral}, {Tice},
  {Trilling}, {Turri}, {Van Berg}, {Vanden Berk}, {Vetter}, {Virieux},
  {Vucina}, {Wahl}, {Walkowicz}, {Walsh}, {Walter}, {Wang}, {Wang}, {Warner},
  {Wiecha}, {Willman}, {Winters}, {Wittman}, {Wolff}, {Wood-Vasey}, {Wu},
  {Xin}, {Yoachim}, \& {Zhan}}]{2019ApJ...873..111I}
{Ivezi{\'c}}, {\v{Z}}., {Kahn}, S.~M., {Tyson}, J.~A., {et~al.} 2019, \apj,
  873, 111

\bibitem[{{Kallivayalil} {et~al.}(2013){Kallivayalil}, {van der Marel},
  {Besla}, {Anderson}, \& {Alcock}}]{Kallivayalil_2013}
{Kallivayalil}, N., {van der Marel}, R.~P., {Besla}, G., {Anderson}, J., \&
  {Alcock}, C. 2013, \apj, 764, 161

\bibitem[{{Kim} {et~al.}(2012){Kim}, {Protopapas}, {Trichas}, {Rowan-Robinson},
  {Khardon}, {Alcock}, \& {Byun}}]{Kim2012}
{Kim}, D.-W., {Protopapas}, P., {Trichas}, M., {et~al.} 2012, \apj, 747, 107

\bibitem[{{Koz{\l}owski} {et~al.}(2012){Koz{\l}owski}, {Kochanek}, {Jacyszyn},
  {Udalski}, {Szyma{\'n}ski}, {Poleski}, {Kubiak}, {Soszy{\'n}ski},
  {Pietrzy{\'n}ski}, {Wyrzykowski}, {Ulaczyk}, \&
  {Pietrukowicz}}]{Kozlowski2012}
{Koz{\l}owski}, S., {Kochanek}, C.~S., {Jacyszyn}, A.~M., {et~al.} 2012, \apj,
  746, 27

\bibitem[{{Mirhosseini} \& {Moniez}(2018)}]{Mirhosseini2018}
{Mirhosseini}, A. \& {Moniez}, M. 2018, A\&A, 618, L4

\bibitem[{Moniez(2010)}]{Moniez2010}
Moniez, M. 2010, General Relativity and Gravitation, 42, 2047–2074

\bibitem[{{Moniez} {et~al.}(2017){Moniez}, {Sajadian}, {Karami}, {Rahvar}, \&
  {Ansari}}]{Moniez2017}
{Moniez}, M., {Sajadian}, S., {Karami}, M., {Rahvar}, S., \& {Ansari}, R. 2017,
  \aap, 604, A124

\bibitem[{{Mr{\'o}z} {et~al.}(2019){Mr{\'o}z}, {Udalski}, {Skowron},
  {Szyma{\'n}ski}, {Soszy{\'n}ski}, {Wyrzykowski}, {Pietrukowicz},
  {Koz{\l}owski}, {Poleski}, {Ulaczyk}, {Rybicki}, \& {Iwanek}}]{Mroz2019}
{Mr{\'o}z}, P., {Udalski}, A., {Skowron}, J., {et~al.} 2019, \apjs, 244, 29

\bibitem[{Mróz {et~al.}(2020)Mróz, Udalski, Szymański, Soszyński,
  Pietrukowicz, Kozłowski, Skowron, Poleski, Ulaczyk, Gromadzki, Rybicki,
  Iwanek, \& Wrona}]{Mroz2020}
Mróz, P., Udalski, A., Szymański, M.~K., {et~al.} 2020, The Astrophysical
  Journal Supplement Series, 249, 16

\bibitem[{{Paczynski}(1986)}]{Paczynski86}
{Paczynski}, B. 1986, \apj, 304, 1

\bibitem[{Paturel {et~al.}(1995)Paturel, Vauglin, Andernach, Garnier,
  Marthinet, Petit, Nella, Bottinelli, Gouguenheim, \& Durand}]{Paturel1995}
Paturel, G., Vauglin, I., Andernach, H., {et~al.} 1995

\bibitem[{Pietrzy{\'{n}}ski {et~al.}(2019)Pietrzy{\'{n}}ski, Graczyk, Gallenne,
  Gieren, Thompson, Pilecki, Karczmarek, G{\'o}rski, Suchomska, Taormina,
  Zgirski, Wielg{\'o}rski, Ko{\l}aczkowski, Konorski, Villanova, Nardetto,
  Kervella, Bresolin, Kudritzki, Storm, Smolec, \& Narloch}]{Pietrzynski2019}
Pietrzy{\'{n}}ski, G., Graczyk, D., Gallenne, A., {et~al.} 2019, Nature, 567,
  200

\bibitem[{{Rahvar}(2015)}]{Rahvar_2015}
{Rahvar}, S. 2015, International Journal of Modern Physics D, 24, 1530020

\bibitem[{{Renault} {et~al.}(1997){Renault}, {Afonso}, {Aubourg}, {Bareyre},
  {Bauer}, {Brehin}, {Coutures}, {Gaucherel}, {Glicenstein}, {Goldman}, {Gros},
  {Hardin}, {de Kat}, {Lachieze-Rey}, {Laurent}, {Lesquoy}, {Magneville},
  {Milsztajn}, {Moscoso}, {Palanque-Delabrouille}, {Queinnec}, {Rich}, {Spiro},
  {Vigroux}, {Zylberajch}, {Ansari}, {Cavalier}, {Couchot}, {Mansoux},
  {Moniez}, {Perdereau}, {Beaulieu}, {Ferlet}, {Grison}, {Vidal-Madjar},
  {Guibert}, {Moreau}, {Maurice}, {Prevot}, {Gry}, {Char}, \&
  {Fernandez}}]{1997A&A...324L..69R}
{Renault}, C., {Afonso}, C., {Aubourg}, E., {et~al.} 1997, \aap, 324, L69

\bibitem[{{Sasaki} {et~al.}(2016){Sasaki}, {Suyama}, {Tanaka}, \&
  {Yokoyama}}]{2016PhRvL.117f1101S}
{Sasaki}, M., {Suyama}, T., {Tanaka}, T., \& {Yokoyama}, S. 2016, \prl, 117,
  061101

\bibitem[{Schneider {et~al.}(2006)Schneider, Kochanek, \&
  Wambsganss}]{Schneider_2006}
Schneider, P., Kochanek, C., \& Wambsganss, J. 2006, Gravitational Lensing:
  Strong, Weak and Micro

\bibitem[{Suntzeff {et~al.}(1988)Suntzeff, Heathcote, Weller, Caldwell, Huchra,
  Olowin, \& Chambers}]{SN1987A}
Suntzeff, N.~B., Heathcote, S., Weller, W.~G., {et~al.} 1988, Nature, 334, 135

\bibitem[{Tisserand(2004)}]{TisserandThese}
Tisserand, P. 2004, PhD thesis

\bibitem[{{Tisserand} {et~al.}(2007){Tisserand}, {Le Guillou}, {Afonso},
  {Albert}, {Andersen}, {Ansari}, {Aubourg}, {Bareyre}, {Beaulieu}, {Charlot},
  {Coutures}, {Ferlet}, {Fouqu{\'e}}, {Glicenstein}, {Goldman}, {Gould},
  {Graff}, {Gros}, {Haissinski}, {Hamadache}, {de Kat}, {Lasserre}, {Lesquoy},
  {Loup}, {Magneville}, {Marquette}, {Maurice}, {Maury}, {Milsztajn}, {Moniez},
  {Palanque-Delabrouille}, {Perdereau}, {Rahal}, {Rich}, {Spiro},
  {Vidal-Madjar}, {Vigroux}, {Zylberajch}, \& {EROS-2
  Collaboration}}]{Tisserand_2007}
{Tisserand}, P., {Le Guillou}, L., {Afonso}, C., {et~al.} 2007, \aap, 469, 387

\bibitem[{Valenti {et~al.}(2015)Valenti, Sand, Stritzinger, Howell, Arcavi,
  McCully, Childress, Hsiao, Contreras, Morrell, Phillips, Gromadzki, Kirshner,
  \& Marion}]{SNIIL}
Valenti, S., Sand, D., Stritzinger, M., {et~al.} 2015, Monthly Notices of the
  Royal Astronomical Society, 448, 2608

\bibitem[{{Whitney} {et~al.}(2008){Whitney}, {Sewilo}, {Indebetouw},
  {Robitaille}, {Meixner}, {Gordon}, {Meade}, {Babler}, {Harris}, {Hora},
  {Bracker}, {Povich}, {Churchwell}, {Engelbracht}, {For}, {Block}, {Misselt},
  {Vijh}, {Leitherer}, {Kawamura}, {Blum}, {Cohen}, {Fukui}, {Mizuno},
  {Mizuno}, {Srinivasan}, {Tielens}, {Volk}, {Bernard}, {Boulanger}, {Frogel},
  {Gallagher}, {Gorjian}, {Kelly}, {Latter}, {Madden}, {Kemper}, {Mould},
  {Nota}, {Oey}, {Olsen}, {Onishi}, {Paladini}, {Panagia}, {Perez-Gonzalez},
  {Reach}, {Shibai}, {Sato}, {Smith}, {Staveley-Smith}, {Ueta}, {Van Dyk},
  {Werner}, {Wolff}, \& {Zaritsky}}]{Whitney2008}
{Whitney}, B.~A., {Sewilo}, M., {Indebetouw}, R., {et~al.} 2008, \aj, 136, 18

\bibitem[{{Wyrzykowski} {et~al.}(2011){Wyrzykowski}, {Skowron}, {Koz{\l}owski},
  {Udalski}, {Szyma{\'n}ski}, {Kubiak}, {Pietrzy{\'n}ski}, {Soszy{\'n}ski},
  {Szewczyk}, {Ulaczyk}, {Poleski}, \& {Tisserand}}]{Wyrzykowski_2011}
{Wyrzykowski}, L., {Skowron}, J., {Koz{\l}owski}, S., {et~al.} 2011, \mnras,
  416, 2949

\end{thebibliography}
\end{document}